\documentclass[prb,twocolumn,superscriptaddress,showpacs,floats,floatfix]{revtex4-1}
\usepackage{dcolumn,amsmath,latexsym,graphicx}
\usepackage[usenames,dvipsnames]{color} \usepackage{multirow}
 \usepackage{array}
 
\RequirePackage{ifpdf} 

\graphicspath{ {./figuras/} } 

\begin{document}

\title{The role of first neighbors geometry in the electronic and mechanical properties of atomic contacts}

\preprint{1}

\author{C. Sabater}
\email{Electronic mail address: carlos.sabater@ua.es}
\altaffiliation[Present address: ]{Chemical Physics Department,
Weizmann Institute of Science,
 76100 Rehovot, Israel.}
\affiliation{Departamento de F\'\i sica Aplicada, Universidad de
Alicante, Campus de San Vicente del Raspeig, E-03690 Alicante,
Spain.}

\author{W. Dednam}
\affiliation{Departamento de F\'\i sica Aplicada, Universidad de
Alicante, Campus de San Vicente del Raspeig, E-03690 Alicante,
Spain.}
\affiliation{Department of Physics, Science Campus, University of South Africa, Private Bag X6, Florida Park 1710, South Africa}

\author{M. R. Calvo}
\affiliation{Departamento de F\'\i sica Aplicada, Universidad de
Alicante, Campus de San Vicente del Raspeig, E-03690 Alicante,
Spain.}
\affiliation{CIC NanoGune, E-20018, Donostia, San Sebastian, Spain}
\affiliation{Ikerbasque, Basque Foundation for Science, 48013 Bilbao, Spain}

\author{M. A. Fern\'andez}
\affiliation{Departamento de F\'\i sica Aplicada, Universidad de
Alicante, Campus de San Vicente del Raspeig, E-03690 Alicante,
Spain.}

\author{C. Untiedt}
\affiliation{Departamento de F\'\i sica Aplicada, Universidad de
Alicante, Campus de San Vicente del Raspeig, E-03690 Alicante,
Spain.}

\author{M. J. Caturla}
\email{Electronic mail address: mj.caturla@ua.es}
\affiliation{Departamento de F\'\i sica Aplicada, Universidad de
Alicante, Campus de San Vicente del Raspeig, E-03690 Alicante,
Spain.}
\date{\today}

\begin{abstract}

We study in detail, by experimental measurements, atomistic simulations and DFT transport calculations, the process of formation and the resulting electronic properties of atomic-sized contacts made of Au, Ag and Cu. Our novel approaches to the data analysis of both experimental results and simulations, lead to a precise relationship between geometry and electronic transmission -- we reestablish the significant influence of the number of first neighbors on the electronic properties of atomic-sized contacts. Our results allow us also to interpret subtle differences between the metals during the process of contact formation as well as the characteristics of the resulting contacts. 

\end{abstract}


\maketitle

\section{Introduction}


Single atoms and molecules have been widely hailed as potential electronic devices over the last twenty years \cite{joachim_electronics_2000}. To make such devices a reality, metallic contact formation and the electrical characteristics of few-atom contacts, need to be understood in depth at the atomic level. The electrical conduction in single-atom contacts has been broadly studied both from an experimental and theoretical point of view \cite{agrait2003quantum}, and single-atom contacts have been proposed as elementary circuit components, such as quantized resistors, capacitors \cite{wang1998capacitance} or switches. \cite{terabe2005quantized}

The conductance of few-atom contacts is given by the sum of contributions from quantized transport modes propagating at the contact junction and 
the number and transmission probabilities of those modes are determined by the size and chemical valence of the central part of the constriction \cite{agrait2003quantum}. For example, both a single-atom contact and a monoatomic chain of Au exhibit a resistance of around a quantum of conductance $G_{0}=2e^2/h$, which is in this case the signature of electronic transport through a single, fully open, quantum channel \cite{Otal13}. 
However, variations in the geometrical configuration of the leads\cite{sabater2013understanding}, i.e., the number of neighboring atoms in the constriction, give rise to fluctuations of up to 20 percent in the conductance of a single atomic contact.
 Not only the electrical properties of single atom contacts are strongly influenced by their coordination to the leads, but also their mechanical properties. 
 When two electrodes in the tunneling regime eventually come into contact, it is known, for certain materials and geometries, that the process of contact formation happens as a sudden jump.  Nonetheless, jump to contact is not a generalized phenomenon and the process of formation may be smooth. \cite{Untiedt_2007} The probability of occurrence of jump to contact and the details of this process have already been suggested to strongly depend not only on the bulk mechanical properties of the material, such as its cohesive energy and Young's modulus\cite{Trouwborst_2008,Fern_ndez_2016},  but also, for certain materials, e.g., Au or Cu, on the specific geometry of the contacting leads \cite{Kr_ger_2009,Kroger16}.

In this article, we focus on the influence of the first-neighbor configurations on the process of formation of single-atom contacts made of Au, Ag and Cu, as well as their associated conductance values.  To this end, we combine atomistic simulations  and quantum transport calculations \cite{pethica1988stability,landman1990atomistic,Bratkos,brandbyge1995conductance,sorensen1998mechanical,dreher2005Aucontacts} with a detailed analysis of experimental results.
We improve the statistical analysis carried out by Untiedt  \textit{et al.} \cite{Untiedt_2007} for Au, and compare our results with those obtained from the atomistic simulations we perform to determine the most likely first-neighbor structures at first contact, and corresponding conductance values we calculate from Density Functional Theory (DFT) methods. \cite{palacios2001fullerene,palacios2002transport,louis2003keldysh} From such a comparison between simulation and experimental results, we can relate the distribution of contact conductances to specific geometries. In agreement with the results published in Refs. \cite{Untiedt_2007,sabater2013understanding} we find the most likely geometries to lie within four classes: monomers, dimers, Double Contacts (D.C) and Triple Contacts (T.C). Furthermore, we identify more specific structures within these classes and more interestingly, find the dispersion in conductance values for each of these classes to be a consequence of the variations in the number of first neighbors.  Our analysis provides a precise assignment of the conductance values reported for these configurations, and remarkably, yields a broader distribution of conductance values for the monomer than in previous works on Au, ultimately explaining previous disagreements between experiments and theory. The reason for this can be traced to a higher dependence of the monomer's conductance on the number of first neighbors. We complete our study by carrying out a similar analysis for Ag and Cu. 


\section{Methods}

\subsection{Experimental methods}

\begin{figure}[htp]
 \centering
 \includegraphics[width=0.5\textwidth]{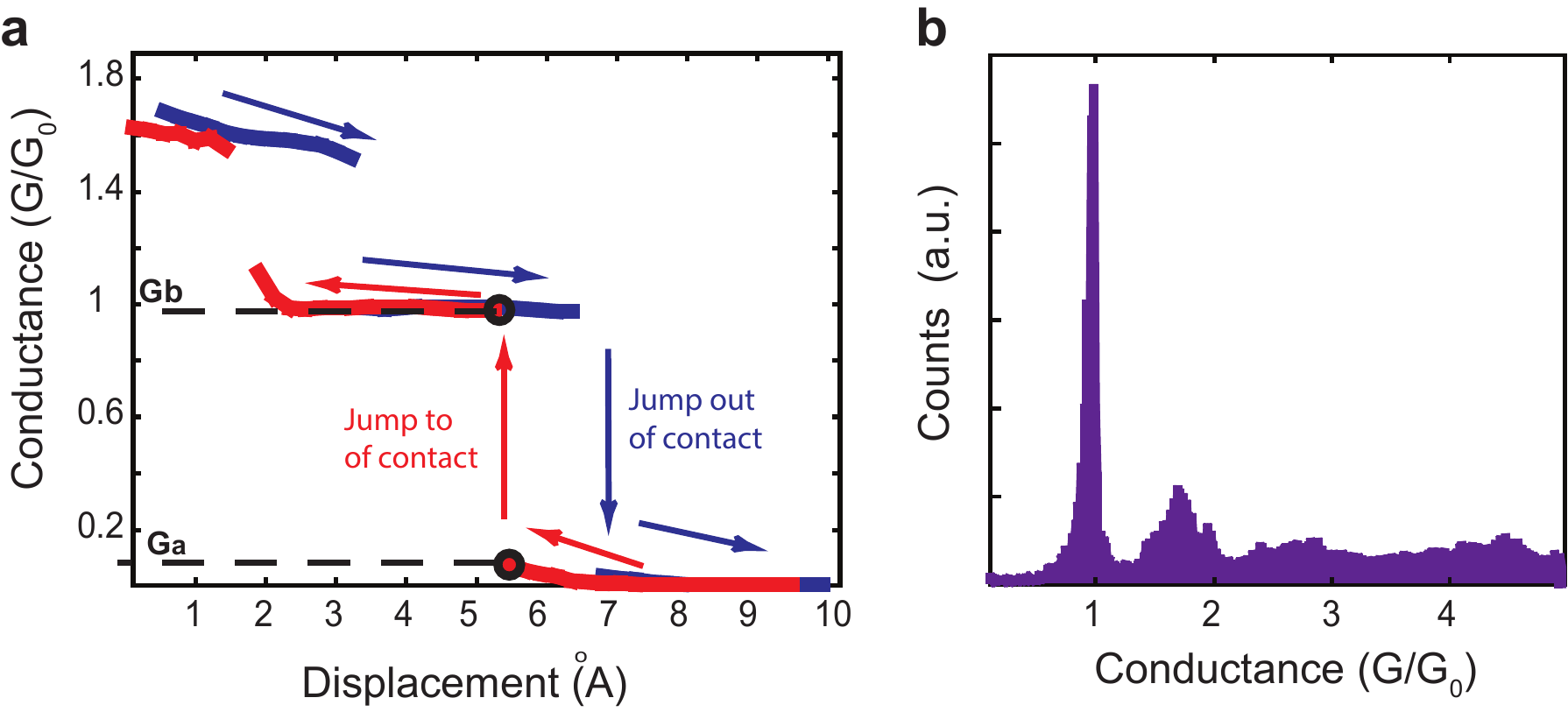}
 \caption{a) Conductance trace for the formation and rupture of a gold atomic contact recorded in our STM-MCBJ setup at 4.2 K. b) Conductance histogram built from more than 1000 Au contact rupture traces.}
 \label{autrhis}
 \end{figure}

Our atomic contacts are fabricated by performing several cycles of indentation and separation of two electrodes made of the same high purity (99.999\%) metal,  Au, Ag or Cu, under cryogenic vacuum at 4.2K.  The electrical conductance of the junctions (obtained as the measured current divided by the applied voltage of 100 mV) is recorded while the two electrodes are carefully brought into contact in a scanning tunneling microscope (STM) setup, as described in previous works. \cite{Untiedt_2007,sabater2013understanding}  The traces of conductance as a function of electrode distance (Fig.\ref{autrhis}(a)) contain valuable information about the process of contact rupture and formation. When electrodes are close enough but not yet in contact, electrons may tunnel between them. In the tunneling regime,  the conductance increases exponentially as the separation between leads decreases. The conductance increases smoothly until a sudden jump occurs, from the tunneling regime up to a clear plateau at around 1 $G_{0}$, indicating  the formation of a monoatomic contact \cite{agrait2003quantum}. Examples of rupture and formation traces are displayed in Fig.1(a).

Every realization of an atomic-size contact produces a slightly different conductance trace, which is suggestive of a variation in structural configurations. Therefore, a statistical analysis of the data is key to extracting information about the most probable configurations. 
An approach that is widely used in the literature \cite{agrait2003quantum} is the construction of a conductance histogram (such as the one in Fig. 1b for the case of rupture traces of Au), to determine the conductance values associated with the most probable configurations of the single-atom contact.

A more specific method for the study of contact formation was introduced by Untiedt et al. \cite{Untiedt_2007} 
As sketched in Fig. 1a, for each formation trace, the highest jump in conductance between two consecutive points is monitored. Two conductance values are then recorded, $G_{a}$, from which the jump occurs and $G_{b}$, the final value immediately after the jump. A density plot of the pairs $(G_{a},G_{b})$ (main panel in Fig. 2) displays the values of greatest probability from and to which the conductance jump occurs.

\begin{figure}[htp]
 \centering
\includegraphics[width=0.45\textwidth]{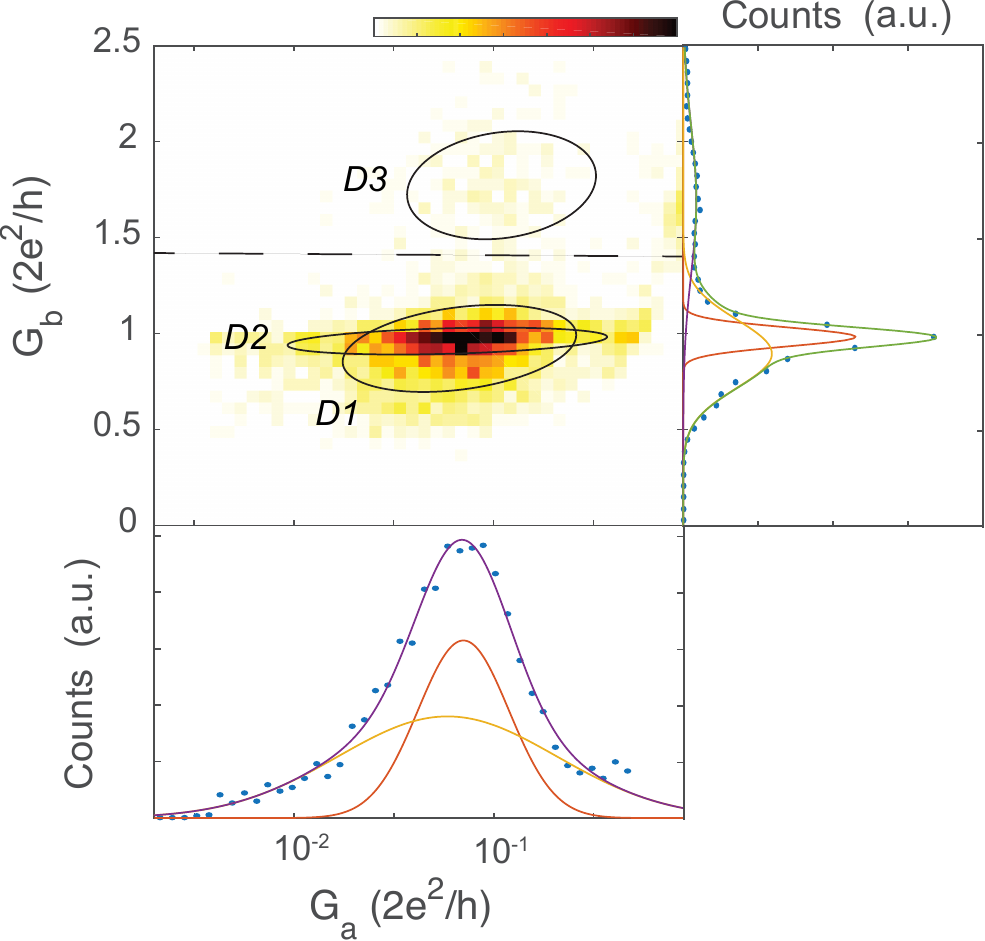}
 \caption{Central panel: density plot constructed from the pairs $(G_a,G_b)$ obtained as described in the text from more than 2000 traces of formation of Au contacts. Right panel: Scatter plot showing the projection of the density plot on the $G_b$ axis. As shown in previous works \cite{Untiedt_2007}, this projection can be fitted to the sum of three gaussian peaks (green line). The purple, orange and yellow lines represent the individual gaussian components. Bottom panel: Projection of density plot on the $G_a$ axis (scatter plot). The maximum above the dashed line has been left out of the analysis here in order to more clearly identify the components of the maximum below the line. The projection of the latter maximum can be fitted to the sum of two gaussian distributions (purple line). The individual components are shown as the yellow and orange lines.}
 \label{Fig2}
 \end{figure}

As mentioned above, prior to contact formation, the tunneling conductance depends exponentially on the distance between electrodes as $G\simeq Ke^{-\frac{\sqrt{2m\phi}}{h}d}$, where 
$K$ is a proportionality constant which depends on the cross-sectional area and density of states at the Fermi level of the electrodes, $m$ corresponds to the electron mass and $\phi$ is the work function of the material.
Since $G_a$ is the conductance in the tunneling regime immediately before jump to contact, its logarithm $log(G_a)$ \footnote{log denotes here the common logarithm (base 10)} is proportional to the distance between the electrodes from which the jump occurs. 
When the $G_a$ axis is plotted on a logarithmic scale, the density plot corresponding to formation of Au contacts, reveals shapes of the maxima that can be more easily interpreted than those previously reported in Ref. \cite{Untiedt_2007} 

\subsection{Data analysis}

The projections of the density plot data on both $log(G_a)$ and $G_b$ axes (Fig. 2) can be fitted to a sum of gaussian peaks. This suggests that the density plot is formed by a number of maxima which are normally distributed in both variables. Therefore, we fit the data to the sum of three bivariate normal distributions, sketched as ellipses in Fig. 2 and labeled $D1$, $D2$ and $D3$, with different relative probabilities $p$. Each of these distributions is described by the expression:
 \begin{equation}
f(x,\mu,\Sigma)=\frac{1}{\sqrt{|\Sigma|}(2\pi)^2}e^{-\frac{1}{2}(x-\mu)'\Sigma^{-1}(x-\mu)}
 \label{eq:xmusigma}
 \end{equation}

 where $x=(log(G_a),G_b)$, $\bf{\mu}=(\mu_a, \mu_b)$ and  $\Sigma=\left( \begin{array}{cc}
 \sigma_a^2 &\rho\sigma_a\sigma_b \\
  \rho\sigma_a\sigma_b & \sigma_b^2   \end{array} \right) $. $\mu_i$ and $\sigma_i$ represent the 2D equivalents of the unidimensional mean and standard deviation, respectively, and $\rho$ is the correlation parameter between variables $logG_a$ and $G_b$. 
  
  \section{Experimental Results}
  The experiments and analysis described in the previous section were repeated during the fabrication of over 2000 contacts made of Au, Ag and Cu. The output fitting parameters for the three materials are summarized in Table I. 
  The characteristic parameters of the distributions can be graphically represented by an ellipse (for example, as the overlays in Fig. 2). The center of the ellipse ($\mu_a$,$\mu_b$) represents the (log$G_a$,$G_b$) position of the mean of the distribution, and the axes of the ellipse represent the standard deviations ($\sigma_a$,$\sigma _b$) in the respective conductance axis. The tilt of the ellipse is proportional to the correlation ($\rho$) between the two variables.
  The identification of three maxima is in good agreement with  Ref. \cite{Untiedt_2007} for Au. This new analysis provides an opportunity to revise those results and carry out a more precise quantitative analysis of the data. 
  
  In analogy with Ref. \cite{Untiedt_2007}, we find an isolated distribution with a low probability of occurrence, well above $G_0$ (labeled $D3$ in Fig. 2), while, at around 1 $G_0$, we find the sum of two distributions. Here we disentangle those two distributions and provide an estimate of their relative probabilities ($p$). Distribution $D1$ contains more than 50 percent of the data, while $D2$ contains around 30 percent. In this instance, the results for all three materials are similar.

\begin{table}
 \centering
\begin{tabular}{|p{1cm}||p{1cm}||p{1cm}|p{1cm}||p{1cm}|p{1cm}|p{1cm}|}
 \hline
 \multicolumn{7}{|c|}{\bf{\textcolor{blue}{Au}}} \\
 \hline
 \centering
   & \centering p (\%) &\centering $ \mu_{a} $&\centering $\mu_{b} $& \centering $\sigma_{a}$& \centering $\sigma_{b}$& \centering\arraybackslash $\rho$\\
 \hline
\centering \bf{\textcolor{blue}{D1}} & \centering 58 &\centering -1.2 &\centering 0.9 &\centering 0.4 &\centering 0.2 &\centering\arraybackslash  0.3 \\
\centering \bf{\textcolor{blue}{D2}} &\centering 32\centering &\centering -1.2 &\centering 1.0 &\centering 0.4 &\centering 0.05 &\centering\arraybackslash  0.3\\
\centering \bf{\textcolor{blue}{D3}} &\centering 10 &\centering -1.1 &\centering 1.7 &\centering 0.4 &\centering 0.3 &\centering\arraybackslash 0.2 \\
 
 \hline
 \hline
\end{tabular}
\\
\begin{tabular}
{|p{1cm}||p{1cm}||p{1cm}|p{1cm}||p{1cm}|p{1cm}|p{1cm}|}
\hline
 \multicolumn{7}{|c|}{\bf{\textcolor{red}{Ag}}} \\
 \hline
& \centering p (\%) &\centering $ \mu_{a} $&\centering $\mu_{b} $& \centering $\sigma_{a}$& \centering $\sigma_{b}$& \centering\arraybackslash $\rho$\\
 \hline
\centering \bf{\textcolor{red}{D1}} &\centering 52 &\centering -0.6 &\centering 1.1 &\centering 0.2 &\centering 0.2 &\centering\arraybackslash  0.4 \\
\centering \bf{\textcolor{red}{D2}} &\centering 30 &\centering -0.9 &\centering 1.0 &\centering 0.2 &\centering 0.08 &\centering\arraybackslash  0.5\\
\centering \bf{\textcolor{red}{D3}} &\centering 18 &\centering -0.6 &\centering 1.9 &\centering 0.2 &\centering 0.2 & \centering\arraybackslash 0.4 \\
 
 \hline
\end{tabular}
\\
\begin{tabular} 
{|p{1cm}||p{1cm}||p{1cm}|p{1cm}||p{1cm}|p{1cm}|p{1cm}|}
 \hline
 \multicolumn{7}{|c|}{\bf{\textcolor{PineGreen}{Cu}}} \\
 \hline
& \centering p (\%) &\centering $ \mu_{a} $&\centering $\mu_{b} $& \centering $\sigma_{a}$& \centering $\sigma_{b}$& \centering\arraybackslash $\rho$\\
 \hline
\centering\bf{\textcolor{PineGreen}{D1}} &\centering 57 & \centering -0.6 &\centering 1.0 &\centering 0.3 &\centering 0.2 &\centering\arraybackslash  0.12 \\
\centering\bf{\textcolor{PineGreen}{D2}} &\centering 29 &\centering -0.8 &\centering 1.0 & \centering 0.3 &\centering 0.08 &\centering\arraybackslash  0.2 \\
\centering\bf{\textcolor{PineGreen}{D3}} &\centering 14 &\centering -0.6 &\centering 1.8 &\centering 0.3 &\centering 0.2 &\centering\arraybackslash  0.2 \\
 \hline
\end{tabular}
\caption{Parameters obtained by fitting density plots of Au, Ag and Cu to the sum of three bivariate normal distributions, namely D1, D2 and D3. $p$ represents the relative probability of each distribution. For simplicity we denote $\mu_log(G_a/G_0)$ and $\sigma _log(G_a/G_0)$ as $\mu_a$ and $\sigma_a$. $\mu_a$ and $\mu_b$ thus represent the mean values of distributions in the log$G_a$ and $G_b$ axis, respectively. $\sigma_a$, $\sigma_b$ are the corresponding standard deviation, and $\rho$ represents the correlation between the two axis. Conductance values are given in quantum units of conductance $G_0=2e^2/h$.}
\end{table}
  
  Moreover, on comparing the three materials, we discover a striking result: there is an important difference between the jump distance of Au versus Ag and Cu, represented by their mean values of log($G_a/G_0$) denoted for simplicity as $\mu_a$. This is the focus of a separate study \cite{PRL}, in which we show that the origin of this phenomenon can be traced to the different strengths of relativistic effects in these materials.
  
  Besides the information given 
  by the mean of each distribution, the standard deviation also provides a measure of the dispersion in each. Regarding the dispersion in the $G_a$ axis (\textit{$\sigma_a$}), which may appear to be much larger in the case of Au, if scaled to the mean value, it is actually similar to the dispersion for Ag and Cu. This indicates, in all cases, that the variation in jump distance is a percentage of the average distance, which, in turn, supports the interpretation provided in Ref. \cite{PRL}, that the dispersion in  conductance originates from the large number of possible geometrical configurations.

While the dispersion in conductances before jump ($\sigma_a$) remain similar in all three distributions for each material, remarkably, the dispersions in $G_{b}$ (\textit{$\sigma_b$}) exhibit significant differences. Distribution $D1$, in contact conductance $G_b$, is rather broad, while distribution $D2$, the second-most probable, exhibits a rather narrower dispersion in this parameter, as is evident from the widths of the ellipses in the $G_b$ axis $\sigma_b$.
This point will be discussed further in light of atomistic simulations, but it already suggests that the conductance in contact of one of the distributions is considerably less sensitive to geometrical variations than the other. 

Finally, we note that the correlation between $G_a$ and $G_b$, \textit{$\rho$} (visible from the tilt of the ellipses) is very similar not only for the three distributions, but also for all three materials, indicating a slight tendency for contacts associated with shorter jump distances to exhibit higher conductances. 

Besides the notable discrepancies in $\mu_a$, a comparison of the metals yields also a number of subtle differences that are connected to the longer jump distance in the case of Au. Firstly, the means $\mu_a$ of $D1$ and $D2$ for Au, occur at about the same distance, while $D3$'s mean has a slightly different value. However, for Ag and Cu, distributions $D1$ and $D3$ are centered at similar values of $log G_a$, while the contacts corresponding to $D2$ are established from a greater jump distance. Regarding the value of \textit{$\mu_{Gb}$}, note the lower conductance value for $D1$ in the case of Au with respect to the other two distributions, as well as with respect to the corresponding values for Ag or Cu.  Although differences between $D1$ and $D2$ are small and perhaps within error margins, this behavior is expected for the more "stretched out" structures formed in Au \cite{PRL}. 


\section{Molecular Dynamic simulations and \textit{Ab-Initio} calculations}
 \subsection{Methodology}
We have not found any analysis of experimental measurements of electronic transport in the literature, which can provide information about the geometry at or the instant just before contact is established. Therefore, in order to have an appreciation of the importance of the configuration of the atoms in the immediate vicinity of few-atom contacts, we simulate the experiments by means of classical molecular dynamics (CMD) and first-principles quantum transport calculations. An alternative approach is used in Refs. \cite{Hybertsen_2016,Hybertsen_2017}, in which a potential energy surface is calculated as an adiabatic trajectory by DFT. Metal junctions composed of small opposing fragments of Au, Ag or Cu are elongated/separated in small steps with a geometry optimization at each step

\begin{figure}[htp]
\centering
 \includegraphics[width=0.48\textwidth]{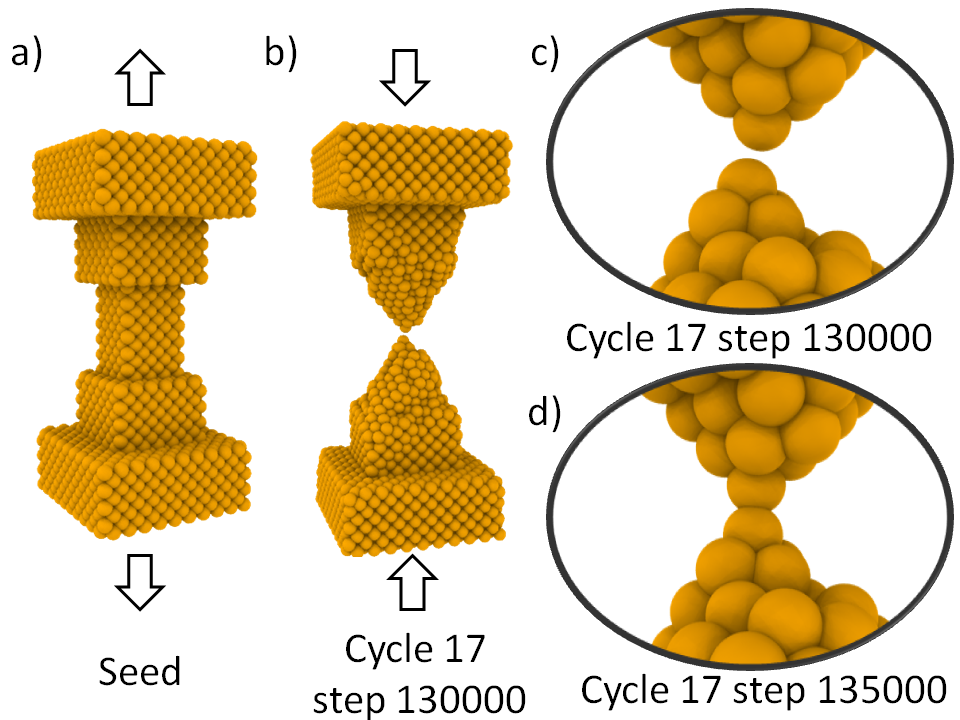}
\caption{Snapshots of a gold nanocontact at different times during a molecular dynamics simulation. Panel: a) initial structure, arrows in a) and b) indicate the direction of elongation or compression. Panel c) shows a zoom-in of panel b), which is the step immediately before the contact shown in panel d) has formed.}
\label{MDfigure}
\end{figure}

Molecular dynamics simulations are based on solving Newton's second law for all the atoms, as they evolve from their initial positions. In such simulations, the potential used to model interactions between the atoms is semi-empirical. \cite{allen1989computer} The initial structure in the present work is independent of the metal and consists of 4736 atoms, oriented along the [100] crystallographic direction, as shown in panel a) of Fig. \ref{MDfigure}. The result of solving Newton's second law for this system is that we can obtain the classical trajectories of all the atoms in the structure, as it is ruptured and brought back into contact over many cycles. Extracting from these trajectories, then, the structure at first contact, as well as the one immediately before it, will, via DFT transport calculations \cite{palacios2001fullerene,palacios2002transport,louis2003keldysh}, yield the conductance at the moment that contact is re-established.

As mentioned above, all the simulations involving Au, Ag and Cu are based on the same initial seed structure. The simulations are run in a way that reproduces cyclic loading of the nanowire in analogy with a typical STM or mechanically controllable break junction (MCBJ) experiment. This is also an approach that was followed in our previous works. \cite{sabater2012mechanical,sabater2013understanding} The interactions between the metal atoms are modeled by the semi-empirical, embedded atom method (EAM) potential. \cite{daw1983semiempirical} All the simulations have been realized by means of the Large-scale Atomic/Molecular Massively Parallel Simulator LAMMPS. \cite{plimpton1995fast,lammps2} The potential parameters used for Au, Ag and Cu in this work, are taken from Ref. \cite{zhou2001atomic} The potential itself is derived in Ref. \cite{wadley2001potential}

Additionally, in order to mimic the conditions of the experiment as closely as possible, we simulate at the boiling point of liquid Helium, $4.2$ K. The Nose-Hoover thermostat \cite{nose1984molecular,hoover1985canonical} serves to maintain the temperature constant during the cycles of retraction and approach of the nanoelectrodes in the simulations. Thermostatting is performed every 1000 simulation time steps, the time interval that is recommended by the developers of LAMMPS.\cite{lammps2}

The atoms that are located in the first three crystallographic planes from the top of the initial seed structure, as well as the corresponding three planes at the bottom, are pinned to their equilibrium bulk lattice positions so as to constrain their relative positions. The remaining atoms respond dynamically to the bulk motion of these ``frozen" planes. Following, the entire structure is stretched lengthwise (vertically) by moving the frozen layers in opposite directions at a constant speed of $\sim$1 m/s. The arrows in Fig. \ref{MDfigure}, panels a) and b), illustrate the directions of the applied forces on both ends (top/bottom) during contact rupture and formation. A speed of $\sim$1 m/s may be many orders of magnitude greater than that employed in the experiments, but we argue that there is enough time for the structures to reach equilibrium, and not merely meta-stable states, because this speed is at least three orders of magnitude lower than that of sound in the bulk metals. \cite{sorensen1998mechanical} The low temperature used in our simulations also ensures that processes that would otherwise be important at microsecond time-scales, such as surface diffusion, remain negligible. In fact, at $4.2$ K, surface diffusion is inhibited by activation energies that are $3-4$ orders of magnitude higher than the thermal energy of the atoms. \cite{ibach2006physics}

To perform cyclic loading in CMD, the simulation domain is divided longitudinally into slices of equal height, corresponding to the interlayer spacing within the bulk crystal. In a face-centered cubic crystal, this spacing is half the lattice parameter along the [100] crystallographic axis. The slice containing the least number of atoms then corresponds to the minimum cross section of the nanocontact. Hence, the structure is stretched until the minimum-atom slice and either of the slices adjacent to it no longer contain any atoms as shown in Fig. \ref{MDfigure} c). At this point, the motion is reversed and the two ruptured tips are brought back together at the same speed with which the structure was first broken. When the minimum-atom layer contains more than 15 atoms, the motion is once more reversed and the nanocontact is stretched until it breaks. This process is repeated at least 20 times. To clarify our terminology, we denote by one ``cycle" a single rupturing and re-forming of the contact.
It is crucial in our simulations to know at which time step during approach, first contact occurs. We detect this moment by monitoring the value of the minimum cross section, which happens when there are more than 0 atoms in the contact cross section. This means that contact has been (re-)established. Incidentally, the semi-empirical potentials describing the interactions between the atoms in the simulations, lead to first-contact distances ranging up to half-way between first and second neighbors in a bulk FCC lattice: $\sim3.5$ \AA \space in the cases of Au and Ag, and $\sim3.0$ \AA \space in the case of Cu. In past works, this has also been used as the criterion to identify the moment of first contact.\cite{Fern_ndez_2016,Dednam2014contacts} Figure \ref{MDfigure} c) and d) show the structure prior to and immediately after first contact, respectively. 

Finally, to calculate the conductance of structures extracted from molecular dynamics simulation trajectories, we have used the electronic transport code ANT.G, \cite{ANTG} which depends on DFT parameters calculated by GAUSSIAN09. \cite{GAUSSIAN09} The structures obtained from CMD contain more than 4000 atoms. Therefore, in order to compute the conductance of these structures within a reasonable time via DFT calculations, it has been necessary to trim the region of interest down to around 500 atoms, keeping only those atoms that lie within a box smaller than the original simulation domain, and centered on the region of first contact, or minimum cross section. However, obtaining accurate conductance values required, in addition, that we had to assign a larger basis set of 11 valence electrons to 40 atoms in the contact region. The rest of the atoms were assigned a basis set of one valence electron.

\subsection{Molecular Dynamics Results\label{SecMD}}

\begin{figure}[htp]
 \centering
 \includegraphics[width=0.51\textwidth]{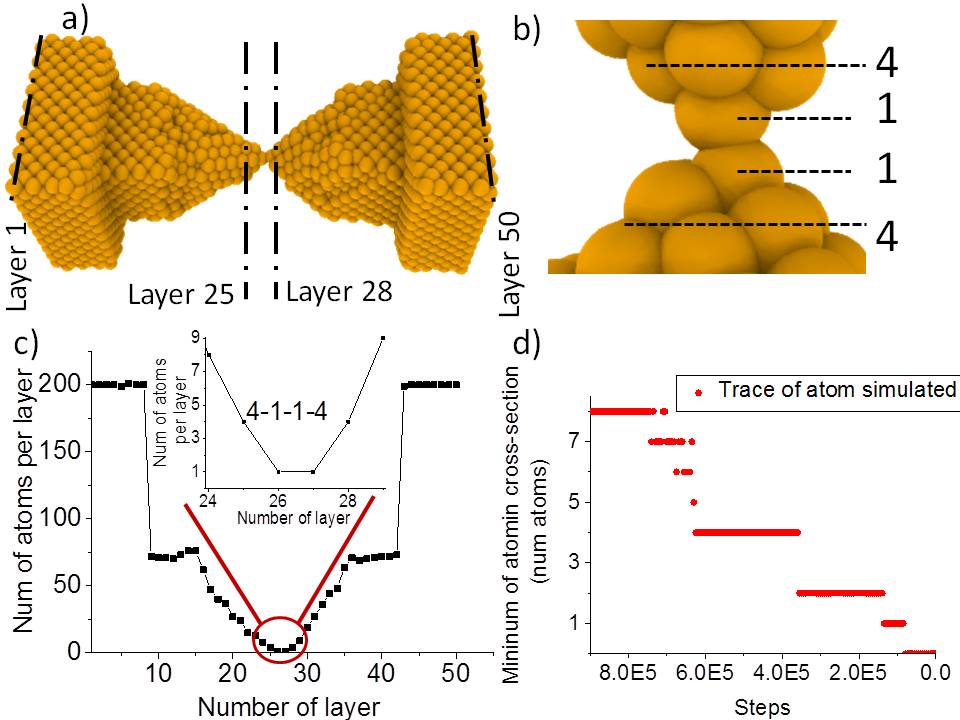}
\caption{a) The atomic-sized gold contact simulated via MD, with the the layer positions indicated by dashed-dotted lines. b) A zoom-in of panel a), showing the type of contact identified by our algorithm, the results of which are shown in panel c). d) The number of atoms in the minimum cross-section as a function of simulation time step. The inset in c) 
is a zoom-in that shows when, during the simulation, exactly 1 atom remains in the minimum cross-section.} 
  \label{figexplacative}
 \end{figure}

For the analysis of the CMD results obtained after 20 cycles of contact rupture and formation, we have used a simple algorithm that counts the number of atoms in layers spaced vertically along the simulation domain. By keeping in mind that the three layers on opposite ends of the structures remain ``frozen" internally during the simulations, i.e., that the lattice parameter of these layers stays fixed at the bulk value, we discretize 
the entire structure into a number of layers half the bulk lattice parameter in thickness. As lattice parameters, we used 4.08 \AA \space for Au and Ag, and 3.61 \AA \space for Cu. Consequently, during an approach (contact formation) phase, for example, we count, at every step, the number of atoms in each layer. 
Figure \ref{figexplacative} a) shows how the layers are distributed along the length of the nanocontact. The plot in Fig. \ref{figexplacative} c) was constructed by counting the number of atoms in each layer. Thus, in principle, a zoom-in of the atoms in the minimum cross section in a), located somewhere between layers 24 and 29, should lead us to conclude that the contact type is ``4-1-1-4". 
Panel b) is such a zoom-in of panel a) and shows clearly what the contact type is. It therefore confirms, via visual inspection, the result inferred from panel c). The trace in Fig. \ref{figexplacative} d) has been constructed by plotting the minimum of the parabola in c) against simulation time step.  The resemblance to an experimental conductance trace is, at the very least, suggestive.
Furthermore, we would like to point out that panel c) contains more information than is used for the purposes of the present article. In fact, such a plot can also give us an idea about the evolution of the sharpness of the contact. For example, blunt electrodes will give rise to broader parabolas than sharper tips. This tool could open the way to a novel analysis of the evolution of the contact in CMD, one that renders direct visualization unnecessary. In addition, a better counting algorithm could take advantage of it. All the results in Fig. \ref{figexplacative} have been extracted from cycle 5 of the simulation involving Au, in which contact occurs at time step 85000.


The methodology followed to count atoms in the cross section is not unique. Other algorithms, such as the one developed by Bratkovsky \textit{et al.} \cite{Bratkos} do not count an integer number of atoms and neighbors in the contact minimum cross section. In this work, we have modified the Bratkovsky algorithm to suit our purposes and count an integer number of atoms in the layers. We are well aware of the limitations of our method, therefore,  
to obtain complementary information, we calculate the conductance of the CMD structures via DFT and if, in the worst of cases, it differs very much from the expected value, we recheck the structure by visual inspection, and where necessary reassign an appropriate contact type.
 
\begin{figure}[htp]
 \centering
 \includegraphics[width=0.49\textwidth] {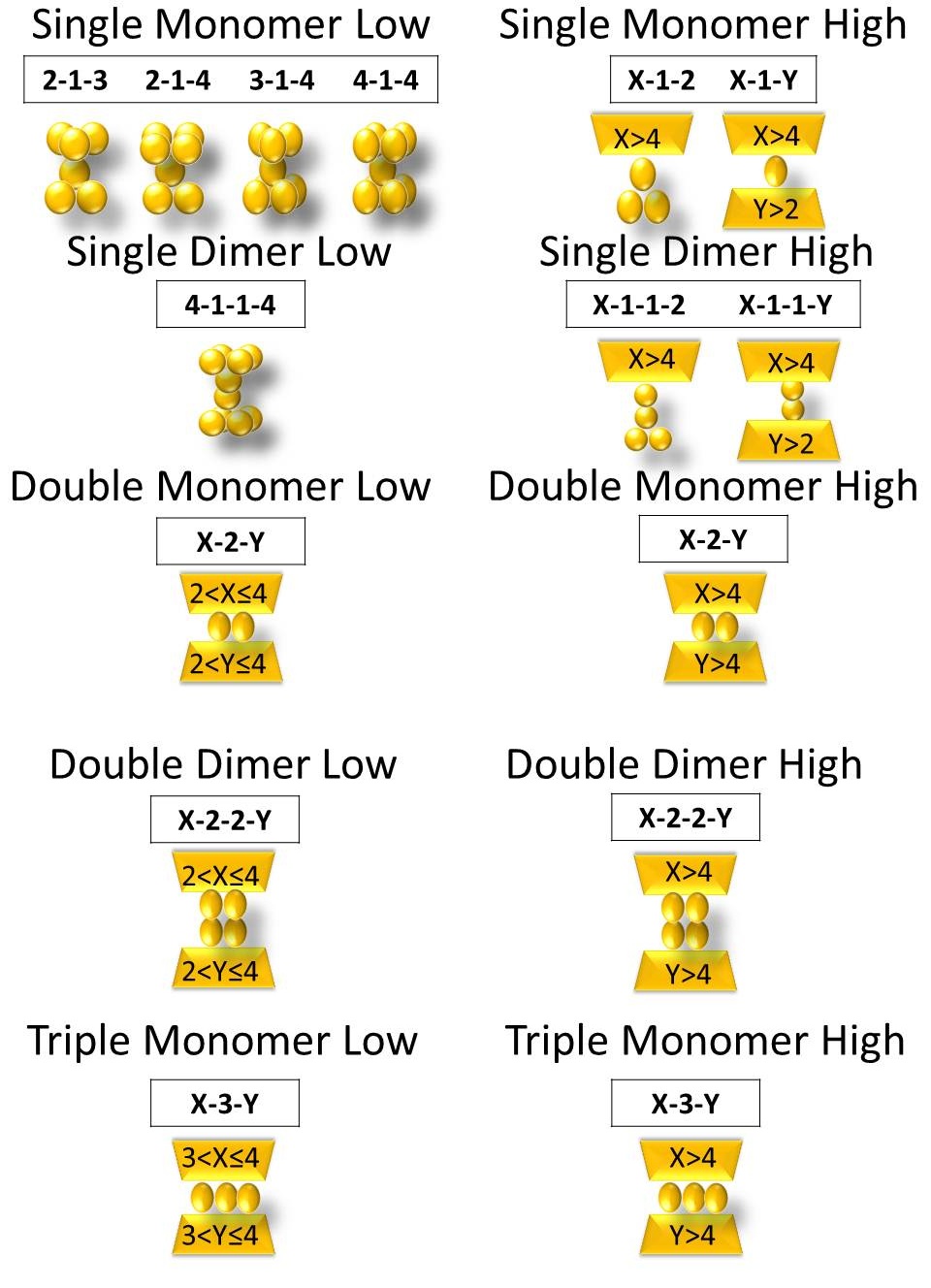}
  \caption{Illustration of the different types of contacts. Left column: low-coordination first-neighbor contacts. Right column: high-coordination first-neighbor contacts. Each of the single, double or triple contacts can also occur as monomers or dimers.}
  \label{Typeofcontact}
  \end{figure}

Thus, we have employed the approach summarized in Fig. \ref{figexplacative}, to study the 3 metals and the 20 cycles of contact rupture-formation they undergo during the simulations. By following the criterion that is outlined in the next paragraph, we have been able to identify different types of contacts as well as their first neighbors, as detailed in Fig.  \ref{Typeofcontact}.
  
Our criterion for identifying the contacts as single, double or triple involves counting the number of atoms in the minimum cross section between the leads, at the very moment when the corresponding layers become populated during the simulation. All three contact types can occur in a monomeric or dimeric configuration, as illustrated in Fig. \ref{Typeofcontact}.  
The ``low" and ``high" coordination designations, irrespective of whether the contacts are monomeric or dimeric, depend on the number of first neighbors found by our algorithm, on either side of the minimum-atom layer. We have established the limit of first neighbors based on an exposed (001) FCC surface layer, which, as is known, is puckered by four-fold hollows, such that an adsorbed atom will have 4 first neighbors immediately beneath it. \cite{Fern_ndez_2016} Then, ``low" coordination means equal to or less than 4 first neighbors, in both electrodes. As soon as the limit of 4 first neighbors is exceeded at one of the electrodes, that side is designated as ``high" coordination. Figure \ref{Typeofcontact} summarizes the typical contacts encountered in our simulations. 

For some of the contacts that form, there is an indeterminate number of possible configurations, and therefore, to simplify the statistical analysis, we use an X to represent combinations with more than 4 first-neighbor atoms. Likewise, we use a Y in combinations where the number of first-neighbor atoms are in a similar range as or larger than X (See Fig. \ref{Typeofcontact}).
 
 Hence, we have simulated contact evolution over continuous loading cycles, and studied the electronic transport during contact formation by means of DFT calculations. After 20 cycles, some of the contact types are reproduced several times, while other contact types appear only once. Table \ref{MDandDFTtab} records, for every cycle, the contact type and number of first-neighbor atoms according to the nomenclature outlined in Fig. \ref{Typeofcontact}. In the same table, we have corrected the type of contact through visual inspection. Raw data about the type of contact, i.e., in the absence of visual inspection, is collected in table \ref{tablaraw}, in the appendix. 
Finally, the double and triple asterisks in table \ref{MDandDFTtab} refer to those curious cases in which 2 or 3  atoms close to forming a contact, contribute to the conductance across the junction, but directly via tunneling.

\subsection{DFT Calculations based on CMD simulations}
    All the MD frames that have been analyzed from the point of view of the geometry in table \ref{tablaraw}, have also been analyzed via DFT conductance calculations. The results are shown in table \ref{MDandDFTtab} and are, in addition, included in Fig. \ref{Projections}. 

\begin{table}[htp]
\centering
 \caption{Conductance values obtained for MD snapshots selected by our first-neighbor visual correction. The colors blue, red and green represent Au, Ag and Cu, in that order. X and Y are any value bigger than 4.}

\begin{tabular}{|c|c|c|c|c|c|c|} 
\hline
\bf{Cycle} & \bf{\textcolor{blue}{Type}} & \bf{\textcolor{blue}{G[$G_(0)$]}}  &  \bf{\textcolor{red}{Type}}  & \bf{\textcolor{red}{G[$G_(0)$]}} & \bf{\textcolor{PineGreen}{Type}} & \bf{\textcolor{PineGreen}{G[$G_(0)$]}}  \\
\hline \hline 

  1 & \textcolor{blue}{4-1-5} &  \textcolor{blue}{1.26}  &\textcolor{red}{6-2-6} & \textcolor{red}{2.43} & \textcolor{PineGreen}{6-3-6}  & \textcolor{PineGreen}{3.25} 
\\2 & \textcolor{blue}{3-1-3} &  \textcolor{blue}{0.72}   &\textcolor{red}{X-2-Y**} & \textcolor{red}{0.69} & \textcolor{PineGreen}{6-3-4}  & \textcolor{PineGreen}{1.71}

\\3 & \textcolor{blue}{X-3-Y} &   \textcolor{blue}{2.45}   &\textcolor{red}{4-1-2} & \textcolor{red}{0.78} & \textcolor{PineGreen}{4-1-3}  & \textcolor{PineGreen}{0.84}

\\4 & \textcolor{blue}{2-1-4} &   \textcolor{blue}{1.58}  &\textcolor{red}{3-1-2} & \textcolor{red}{0.95} & \textcolor{PineGreen}{8-3-6}  & \textcolor{PineGreen}{2.31} 

\\5 & \textcolor{blue}{4-1-1-5} &  \textcolor{blue}{0.73}  &\textcolor{red}{6-2-4} & \textcolor{red}{1.55} & \textcolor{PineGreen}{5-4-8}  & \textcolor{PineGreen}{2.54} 

\\6 & \textcolor{blue}{X-3-Y} &  \textcolor{blue}{2.76} &  \textcolor{red}{8-2-5} & \textcolor{red}{1.72} & \textcolor{PineGreen}{8-3-4}  & \textcolor{PineGreen}{1.58} 

\\7 & \textcolor{blue}{4-3-Y} &  \textcolor{blue}{2.80}  &\textcolor{red}{8-2-4} & \textcolor{red}{1.72} & \textcolor{PineGreen}{5-1-2}  & \textcolor{PineGreen}{1.04} 

\\8 & \textcolor{blue}{4-1-2} & \textcolor{blue}{0.84} & \textcolor{red}{5-1-5} & \textcolor{red}{1.14} & \textcolor{PineGreen}{6-3-4}  & \textcolor{PineGreen}{2.59} 

\\9 & \textcolor{blue}{7-3-4} & \textcolor{blue}{2.34}  &\textcolor{red}{6-2-6} & \textcolor{red}{1.80} & \textcolor{PineGreen}{5-1-4}  & \textcolor{PineGreen}{0.58} 

\\10 & \textcolor{blue}{3-1-4} &   \textcolor{blue}{1.45}  &\textcolor{red}{6-2-2-6} & \textcolor{red}{1.47} & \textcolor{PineGreen}{4-1-1-6}  & \textcolor{PineGreen}{0.83} 

\\11 & \textcolor{blue}{X-2-Y} &   \textcolor{blue}{2.02} &\textcolor{red}{6-2-4} & \textcolor{red}{1.85} & \textcolor{PineGreen}{6-2-4}  & \textcolor{PineGreen}{1.29} 

\\12 & \textcolor{blue}{3-1-2} &   \textcolor{blue}{1.25}  &\textcolor{red}{6-2-4} & \textcolor{red}{1.72} & \textcolor{PineGreen}{5-4-6}  & \textcolor{PineGreen}{3.14}  

\\13 & \textcolor{blue}{X-2-Y} &   \textcolor{blue}{2.22}  &\textcolor{red}{6-2-6} & \textcolor{red}{1.71} & \textcolor{PineGreen}{4-3-5}  & \textcolor{PineGreen}{2.47}   

\\14 & \textcolor{blue}{5-1-1-3} &   \textcolor{blue}{0.86}  &\textcolor{red}{5-1-1-5} & \textcolor{red}{0.99} & \textcolor{PineGreen}{$***$}  & \textcolor{PineGreen}{1.32}  

\\15 & \textcolor{blue}{4-2-3} &   \textcolor{blue}{1.15}  &\textcolor{red}{5-1-1-5} & \textcolor{red}{0.75} & \textcolor{PineGreen}{4-2-4}  & \textcolor{PineGreen}{2.02}   

\\16 & \textcolor{blue}{4-1-1-5} &   \textcolor{blue}{1.27}  &\textcolor{red}{5-1-1-5} & \textcolor{red}{0.61} & \textcolor{PineGreen}{4-1-2}  & \textcolor{PineGreen}{1.00}  

\\17 & \textcolor{blue}{4-1-1-4} &   \textcolor{blue}{0.88}  &\textcolor{red}{6-2-6} & \textcolor{red}{1.66} & \textcolor{PineGreen}{4-3-7}  & \textcolor{PineGreen}{2.34}  

\\18 & \textcolor{blue}{2-1-4} &   \textcolor{blue}{1.63}  &\textcolor{red}{6-2-6} & \textcolor{red}{1.72} & \textcolor{PineGreen}{4-1-2}  & \textcolor{PineGreen}{1.14}   

\\19 & \textcolor{blue}{3-1-3} &   \textcolor{blue}{1.35}  &\textcolor{red}{6-2-6} & \textcolor{red}{1.82} & \textcolor{PineGreen}{8-3-5}  & \textcolor{PineGreen}{1.47}

\\20 & \textcolor{blue}{4-2-6} &   \textcolor{blue}{2.20}  & \textcolor{red}{6-2-6} & \textcolor{red}{1.75} & \textcolor{PineGreen}{X-3-Y}  & \textcolor{PineGreen}{1.72}
\\
\hline

\end{tabular}
\label{MDandDFTtab}
\end{table}

	The structures obtained from CMD simulations, which are limited in their ability to predict realistic structures, require interpretation via electronic transport calculations (if meaningful comparisons with the experimental results are to be made). Following this, upon comparing the calculated conductance and experimental density plots, we can extract information about the type of contact that is formed as well as the configuration of the first-neighbor atoms around it. The electronic transport across all the structures has been calculated by means of ANT.G \cite{ANTG}, which interfaces with GAUSSIAN09. \cite{GAUSSIAN09} 
    We have grouped the various contacts by type, and their mean conductance values and 
    standard deviations are plotted in Fig. \ref{Projections} as dots and vertical bars, respectively.

\section{Discussion}

In this work, our aim is to find the origin of the subtle differences between materials, and identify the properties of types of contacts defined by their specific geometry. Elsewhere, we prove that relativistic effects are responsible for the large discrepancy between the jump-to-contact distances of Au and Ag \cite{PRL}, represented by the respective means of their $G_a$ values.
 
 
  To approach this problem, we use CMD as a tool to visualize the moment of first contact and identify the number and arrangement of the first neighbors. We cannot rely on CMD in the case of tunneling because the potentials only account indirectly for the effects of electrons, and hence, relativity, and, then, only very crudely. 
Furthermore, it is not possible, experimentally, to know the structure and geometry of the electrodes in the tunneling regime. In CMD, the structure before contact is, at times, preserved in contact, as illustrated in Figs. \ref{MDfigure} c) and d). At other times, significant rearrangements occur and the before-contact structures are no longer preserved.
For this reason, we confine our analysis to the first neighbors in the contact regime.

\begin{table}[ht]
\centering
\caption{The first column refers to the type of contact that has been formed: single, double or triple. In the second column, each of these contact types are further classified into monomer or dimer. The ``L" and ``H" designations under the column heading ``Coord" refer to Low and High coordinations, respectively. Then, the following three columns show the  average conductance values and their standard deviations for Gold, Silver and Copper, respectively. Configurations not found in Molecular Dynamics Simulations are indicated by a ``dash".}

\label{Tablesimply}
\begin{tabular}{|c|c|c|c|c|c|}
\hline
\textbf{Type}& \textbf{\begin{tabular}[c]{@{}l@{}} Mon\\ or dim\end{tabular}} & \textbf{Coord} & \bf{\textcolor{blue}{Au}} & \bf{\textcolor{red}{Ag}} & \bf{\textcolor{PineGreen}{Cu}} \\ \hline \hline

\multirow{4}{*}{Single}                                            & \multirow{2}{*}{Mon.}                                              & L                &  \textcolor{blue}{1.4$\pm$0.3}           &       \textcolor{red}{0.85$\pm$0.12}       & \textcolor{PineGreen}{0.99$\pm$0.15}             \\ \cline{3-6} 
                                                                   &                                                                         & \textbf{H}                 &       \bf{\textcolor{blue}{1.26}}      &    \bf{\textcolor{red}{1.14}}          &      \bf{\textcolor{PineGreen}{0.8$\pm$0.3}}        \\ \cline{2-6} 
                                                                   & \multirow{2}{*}{Dim.}                                                & L                &        \textcolor{blue}{0.88}         &         \textcolor{red}{-}        &          \textcolor{PineGreen}{-}       \\ \cline{3-6} 
                                                                   &                                                                         & \textbf{H}               &     \bf{\textcolor{blue}{1.0$\pm$0.2}}          & \bf{\textcolor{red}{0.78$\pm$0.19}}                    & \bf{\textcolor{PineGreen}{0.83}}                     \\ \hline \hline

\multirow{4}{*}{Double}                                             & \multirow{2}{*}{Mon.}                                              & L                  &  \textcolor{blue}{1.15}                   &  \textcolor{red}{-}                   &        \textcolor{PineGreen}{2.02}       \\ \cline{3-6} 
                                                                   &                                                                         &\textbf{H}               &     \bf{\textcolor{blue}{2.21$\pm$0.02}}            &  \bf{\textcolor{red}{1.73$\pm$0.08}}             &         \bf{\textcolor{PineGreen}{1.29}}      \\ \cline{2-6} 
                                                                   & \multirow{2}{*}{Dim.}                                                & L                  & \textcolor{blue}{-}                   &    \textcolor{red}{-}         &       \textcolor{PineGreen}{-}      \\ \cline{3-6} 
                                                                   &                                                                         &\textbf{H   }              &    \textcolor{blue}{-}         &     \bf{\textcolor{red}{1.47}}        &    \bf{\textcolor{PineGreen}{-}}           \\ \hline \hline
\multirow{2}{*}{Triple}                                            & \multirow{2}{*}{Mon.}                                              & L                  &       \textcolor{blue}{-}      &      \textcolor{red}{-}          &         \textcolor{PineGreen}{-}    \\ \cline{3-6} 
                                                                                                                                                   &                                                                         & \textbf{H }                &     \textbf{\textcolor{blue}{2.5$\pm$0.4}}        &    \textbf{\textcolor{red}{-}}        &      \bf{\textcolor{PineGreen}{2.2$\pm$0.6}}         \\ \hline
\end{tabular}
\end{table}

  \begin{figure}[htp]
 \centering
 \includegraphics[width=0.5\textwidth] {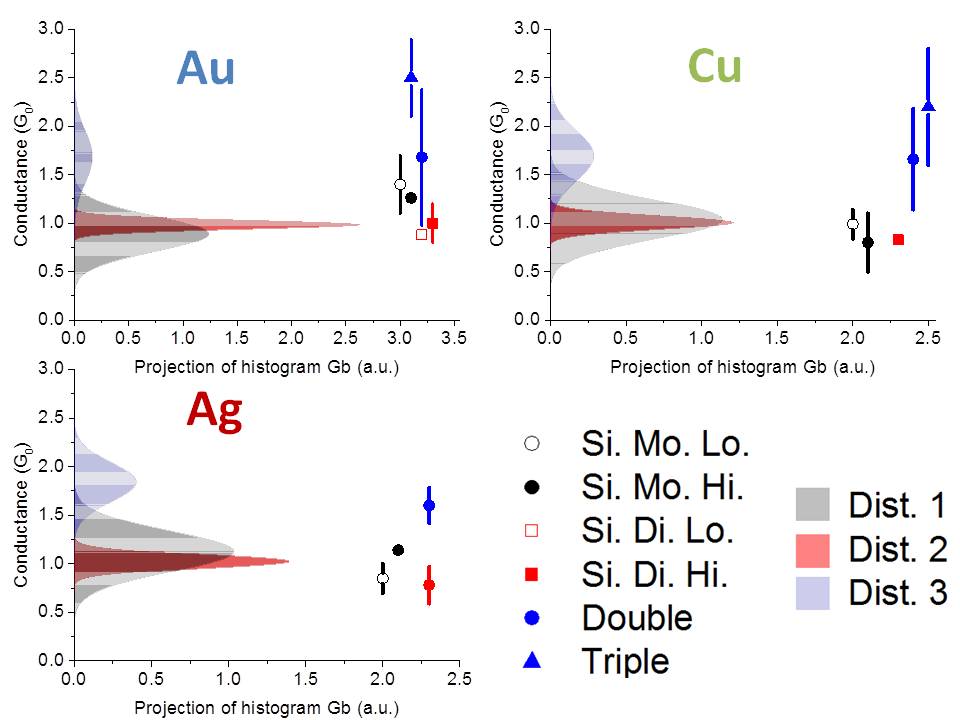}
  \caption{Projection of experimental $G_b$ values vs number of counts, for Au Ag and Cu. Data points and error bars: the conductance and standard deviation of the various simulated contacts. ``Single" is denoted by Si, followed by Mo or Di, depending on whether ``monomer" or ``dimer". Hi and Lo represent high and low coordination}
  \label{Projections}
  \end{figure}
  
The conductance values obtained via DFT from the CMD structures are summarized in table \ref{Tablesimply}. The comparison of these results with the experimental distribution of values (Figure
\ref{Projections}) allows us to interpret our results in terms of the simulated geometry of the contacts. Double and triple contacts are simplified in Fig.  
\ref{Projections}, i.e., we don't distinguish between high or low, or monomer or dimer. Thus, the blue dot and triangle represent mean values, and their error bars, the standard deviations obtained through grouping.


In spite of the reduced statistics (we have performed 20 loading cycles in CMD, on each metal), we observe how the distribution of conductance for the calculated geometries, classified as monomer, dimer and higher order contacts, mostly coincide with the three distributions obtained from the experimental data.  We can therefore confirm the assignment by Untiedt \textit{et al.} \cite{Untiedt_2007}, of distributions $D1$ as monomer, $D2$ as dimer, and $D3$ as higher coordination. 
Our new simulations allow us to further classify the contacts 
into high- and low-coordination. 
This classification does not provide much additional interpretation of the experimental results due to the reduced statistics, but it does highlight the determining role of coordination on the conductance of atomic contacts.

Our results for Au and Cu display a higher dispersion in conductance for monomers ($D1$) than dimers ($D2$)
(Table \ref{Tablesimply}, Fig. \ref{Projections}). The values listed in Table II exemplify how variations on the number of neighbors, for a dimer, have little repercussion on the value of the conductance. For a monomer, on the other hand, the number of neighbors result in large changes in conductance.


We also find that distribution $D3$ likely arises from a combination of double- and triple-contact structures. This leads to a wider distribution in conductance values, as can be seen also from the experimental data. Among these structures, we have identified, through conductance calculations, the triple contact, whose conductance values are in the $2-3$ $G_0$ range. In any event, there may be other structures that have not yet been identified, but that could be discovered by means of the new analysis methods introduced in this work.

Another important difference in the conductance values obtained from the simulations, is the small dispersion in $G_b$ of the monomeric and dimeric Cu structures as compared to Au. As alluded to earlier, in Cu, the dispersion in calculated $G_b$ of the monomer is twice that of the dimer, which is in agreement with the broader $D1$ profile relative to $D2$ in the experimental projections. In Au, the (experimental) $D2$ profile exhibits a very narrow distribution, similar to the narrow dispersion in values obtained for the low-coordinated single dimers from the simulations. This may suggest that these are actually the predominant structures occurring experimentally. 


Finally, in the case of Au, we found a particularly good match between experimental and calculated means and standard deviations, particularly for the dimer, while, for the monomer, the calculated means are slightly over estimated.  Since the simulations do not accurately capture the jump to contact, contact distances are probably shorter, leading to higher expected conductances.

\section{SUMMARY}
 By introducing a new statistical approach that permits identifying properties of atomic-sized contacts with greater precision, it has been possible to study, in detail, the process of formation of Au, Ag and Cu nanocontacts. This analysis allow us to identify with higher precision the distribution of values of conductance associated to different geometries, but also to extract information on the distance of contact formation for those geometries.
  Furthermore, we have used molecular dynamics to simulate the formation of atomic-sized contacts in STM/MCBJ experiments. These simulated contacts were, in turn, analyzed by means of a novel methodology that permits classifying their type and finding the number of first-neighbor atoms in their immediate vicinity. DFT transport calculations on the simulated structures provided a means of comparing theoretical results with the experimental data. We have demonstrated that the type of contact and the geometry of its first neighbors (shape, distance between first-neighbor atoms, and between them and the atomic contact itself) play decisive roles in electronic transport across the simulated contacts. Through a combination of the above three methods, we have found that the electronic transport across the atomic-sized contacts depends crucially on the number of first-neighbor atoms.

\section{ACKNOWLEDGMENTS}

This work has been funded from the Spanish MEC through grants FIS2013-47328 and MAT2016-78625. C.S. gratefully acknowledges financial support from SEPE Servicio P\'ublico de Empleo Estatal. W.D. acknowledges funding from the National Research Foundation of South Africa through the Innovation Doctoral scholarship programme, Grant UID 102574. W.D. also thanks J. Fernandez-Rossier and J.J. Palacios for fruitful discussions.
\section{APPENDIX}

The methodology described in section \ref{SecMD} and illustrated in Fig. \ref{figexplacative} has been applied to the three materials during 20 MD rupture-formation cycles. Table \ref{tablaraw} summarizes the obtained results. It records, for Au, Ag and Cu (in blue, red and green, respectively), the time step (in kilosteps, or, more precisely, picoseconds) when contact is established as well as the type of first contact that is formed during every cycle. Data marked with asterisks indicate that the algorithm has detected a contact when it has not really occurred. Through visual inspection we have selected the correct CMD timeframe in which contact actually occurred and also identified the type of contact. 
 
\begin{table}[htp]
\centering
\caption{Results produced by  modified Bratkovsky algorithm to count integer number of atoms. The colors refer to different materials: Au (blue), Ag (red), and Cu (green).}
\begin{tabular}{|c|c|c|c|c|c|c|} 
\hline
\bf{Cycle}& \bf{\textcolor{blue}{kStep}}  & \bf{\textcolor{blue}{Type}}& \bf{\textcolor{red}{kStep}}  & \bf{\textcolor{red}{Type}} & \bf{\textcolor{PineGreen}{kStep}} & \bf{\textcolor{PineGreen}{Type}} \\
\hline \hline 

  1 & \textcolor{blue}{3090} &  \textcolor{blue}{10-1-2}  &\textcolor{red}{1545} & \textcolor{red}{5-2-5} & \textcolor{PineGreen}{715}  & \textcolor{PineGreen}{6-3-8} 

\\2 & \textcolor{blue}{145} &  \textcolor{blue}{4-1-2}   &\textcolor{red}{1410*} & \textcolor{red}{X-2-Y*} & \textcolor{PineGreen}{540}  & \textcolor{PineGreen}{6-3-4}

\\3 & \textcolor{blue}{285} &   \textcolor{blue}{3-2-5}   &\textcolor{red}{405} & \textcolor{red}{6-1-2} & \textcolor{PineGreen}{220*}  & \textcolor{PineGreen}{4-1-3*}

\\4 & \textcolor{blue}{605} &   \textcolor{blue}{2-1-4}  &\textcolor{red}{395} & \textcolor{red}{3-1-2} & \textcolor{PineGreen}{1055}  & \textcolor{PineGreen}{2-1-6} 

\\5 & \textcolor{blue}{85} &  \textcolor{blue}{4-1-1-4}  &\textcolor{red}{250} & \textcolor{red}{6-2-6} & \textcolor{PineGreen}{335}  & \textcolor{PineGreen}{5-2-3} 

\\6 & \textcolor{blue}{58} &  \textcolor{blue}{4-3-11} &  \textcolor{red}{210} & \textcolor{red}{5-3-5} & \textcolor{PineGreen}{185}  & \textcolor{PineGreen}{4-1-3} 

\\7 & \textcolor{blue}{275*} &  \textcolor{blue}{4-3-Y*}  &\textcolor{red}{320} & \textcolor{red}{6-3-5} & \textcolor{PineGreen}{610}  & \textcolor{PineGreen}{6-2-2-7} 

\\8 & \textcolor{blue}{160} & \textcolor{blue}{6-1-2} & \textcolor{red}{210} & \textcolor{red}{6-2-6} & \textcolor{PineGreen}{285}  & \textcolor{PineGreen}{3-1-4} 

\\9 & \textcolor{blue}{425} & \textcolor{blue}{7-3-4}  &\textcolor{red}{255} & \textcolor{red}{6-3-5} & \textcolor{PineGreen}{165*}  & \textcolor{PineGreen}{5-1-4*}

\\10 & \textcolor{blue}{240} &   \textcolor{blue}{2-1-4}  &\textcolor{red}{315} & \textcolor{red}{6-3-5} & \textcolor{PineGreen}{525}  & \textcolor{PineGreen}{4-1-4} 

\\11 & \textcolor{blue}{400} &   \textcolor{blue}{4-2-5} &\textcolor{red}{255*} & \textcolor{red}{6-2-4*} & \textcolor{PineGreen}{385}  
& \textcolor{PineGreen}{3-2-4} 

\\12 & \textcolor{blue}{435*} &   \textcolor{blue}{3-1-2*}  &\textcolor{red}{320} & \textcolor{red}{5-3-5} & \textcolor{PineGreen}{1060}  & \textcolor{PineGreen}{3-1-6}  

\\13 & \textcolor{blue}{375} &   \textcolor{blue}{7-2-3}  &\textcolor{red}{255} & \textcolor{red}{6-3-5} & \textcolor{PineGreen}{720*}  & \textcolor{PineGreen}{4-3-5*}   

\\14 & \textcolor{blue}{365} &   \textcolor{blue}{7-1-1-2}  &\textcolor{red}{240} & \textcolor{red}{6-2-6} & \textcolor{PineGreen}{1260*}  & \textcolor{PineGreen}{X-4-X*}  

\\15 & \textcolor{blue}{200} &   \textcolor{blue}{5-2-3}  &\textcolor{red}{280} & \textcolor{red}{4-1-6} & \textcolor{PineGreen}{180}  & \textcolor{PineGreen}{8-2-2-4}   

\\16 & \textcolor{blue}{185} &   \textcolor{blue}{2-1-3}  &\textcolor{red}{305} & \textcolor{red}{6-2-6} & \textcolor{PineGreen}{225}  & \textcolor{PineGreen}{5-1-1-3}  

\\17 & \textcolor{blue}{135} &   \textcolor{blue}{4-1-1-4}  &\textcolor{red}{210} & \textcolor{red}{6-3-5} & \textcolor{PineGreen}{225*}  & \textcolor{PineGreen}{4-3-7*}  

\\18 & \textcolor{blue}{535} &   \textcolor{blue}{3-1-2}  &\textcolor{red}{210} & \textcolor{red}{6-3-5} & \textcolor{PineGreen}{265}  & \textcolor{PineGreen}{3-1-1-6}   

\\19 & \textcolor{blue}{340} &   \textcolor{blue}{3-1-2}  &\textcolor{red}{355} & \textcolor{red}{6-3-5} & \textcolor{PineGreen}{410}  & \textcolor{PineGreen}{8-3-5}
 \\20 & \textcolor{blue}{700} &   \textcolor{blue}{4-2-6}  & \textcolor{red}{215} & \textcolor{red}{6-3-5} & \textcolor{PineGreen}{160}  & \textcolor{PineGreen}{8-3-4}
\\
\hline

\end{tabular}

\label{tablaraw}
\end{table}

\bibliography{j2c_relativistic.bib}
  \end{document}